# Accelerator experiments with soft protons and hyper-velocity dust particles: application to ongoing projects of future X-ray missions

E. Perinati*[a], S. Diebold[a], E. Kendziorra[a], A. Santangelo[a], C. Tenzer[a], J. Jochum[b], S. Bugiel[c], R. Srama[c], E. Del Monte[d], M. Feroci[d], A. Rubini[d], A. Rachevski[e], G. Zampa[e], N. Zampa[e], I. Rashevskaya[e], A. Vacchi[e], P. Azzarello[f], E. Bozzo[f], J.-W. den Herder[g], S. Zane[h], S. Brandt[i], M. Hernanz[l], M. A. Leutenegger[m], R. L. Kelley[m], C. A. Kilbourne[m], N. Meidinger[n,q], L. Strüder[n,q], B. Cordier[o], D. Götz[o], G. W. Fraser[p], J. P. Osborne[p], K. Dennerl[q], M. Freyberg[q] and P. Friedrich[q]

[a]IAAT – Institut für Astronomie und Astrophysik, Universität Tübingen, 72076 Tübingen, Germany
[b]Physikalisches Institut, Universität Tübingen, 72076 Tübingen, Germany
[c]MPIK-Max Planck Institut für Kernphysik, 69117 Heidelberg, Germany
[d]INAF - Istituto di Astrofisica e Planetologia Spaziali di Roma, 00133 Roma, Italy
[e]INFN-Istituto Nazionale di Fisica Nucleare-Sez. Trieste, 34127 Trieste, Italy
[f]ISDC-University of Geneve, CH -1211 Geneve, Switzerland
[g]SRON-Space Research Organization Netherlands, 3584 Utrecht, Netherlands
[h]UCL-Mullard Space Science Laboratory, RH5-6NT Dorking, UK
[i]DTU-National Space Institut, 2800 Lyngby, Denmark
[l]CSIS-IEEC-Institut d´Estudis Espacials de Catalunya, 08034 Barcelona, Spain
[m]NASA-Goddard Space Flight Center, 20771 Greenbelt-MD, USA
[n]MPI -Halbleiterlabor, 81739 München, Germany
[o]CEA – Service d'Astrophysique, 91191 Gif-sur-Yvette, France
[p]Department of Physics and Astronomy, University of Leicester, LE1-7RH Leicester, UK
[q]MPE -Max Planck Institut für Extraterrestrische Physik, 85748 Garching, Germany

## ABSTRACT

We report on our activities, currently in progress, aimed at performing accelerator experiments with soft protons and hyper-velocity dust particles. They include tests of different types of X-ray detectors and related components (such as filters) and measurements of scattering of soft protons and hyper-velocity dust particles off X-ray mirror shells. These activities have been identified as a goal in the context of a number of ongoing space projects in order to assess the risk posed by environmental radiation and dust and qualify the adopted instrumentation with respect to possible damage or performance degradation. In this paper we focus on tests for the *Silicon Drift Detectors* (SDDs) used aboard the *LOFT* space mission. We use the Van de Graaff accelerators at the University of Tübingen and at the Max Planck Institute for Nuclear Physics (MPIK) in Heidelberg, for soft proton and hyper-velocity dust tests respectively. We present the experimental set-up adopted to perform the tests, status of the activities and some very preliminary results achieved at present time.

## 1. INTRODUCTION

Silicon-based detectors are largely used in space astrophysics. *Charge Coupled Devices* (CCDs) have been succesfully adopted in the focal plane of missions such as *Chandra* and *XMM-Newton*, providing both good spectral resolution and imaging. New types of Silicon detectors, that promise even improved performance, are envisioned for future missions.

---

*Email: Emanuele.Perinati@uni-tuebingen.de   Tel.: +49 7071 29 73457

*Silicon Drift Detectors* (SDDs) are among these, and they have been selected as the baseline instrumentation for the LOFT space mission. However, it is known that the performance of Silicon detectors in space can be strongly affected by the radiation environment. Hyper-velocity micro-meteoroids and debris also represent a risk. In this work we present the characterization of some properties of SDDs with respect to soft proton irradiation and hyper-velocity particle impacts. Despite the fact that the flux of primary cosmic rays in an equatorial Low Earth Orbit (LEO) is made relatively low by the geo-magnetic shielding, at very soft energies the flux of protons is enhanced by a near-equatorial component originated from the charge exchange mechanism, and the design of the *LOFT* mission suggests that the effects of the dose sustained by the SDDs in orbit are carefully explored. For the same reason, it is also important to verify the robustness of the SDDs against impacts of hyper-velocity micro-meteoroids and debris populating the LEO environment. In Section 2 we shortly review the structure and working principle and we describe how the SDDs will be assembled on *LOFT*. In Section 3 we describe the radiation and dust environments in LEO orbit. In Section 4 we discuss which effects can be expected on SDDs from soft proton irradiation and micro-meteoroid/debris impacts. In Section 5 we present the experimental-set up for soft proton irradiations and some very preliminary analysis on the data recently obtained in a test campaign performed in June 2012. In Section 6 we present the experimental set-up for the micro-meteoroid/debris tests, and some results from a test campaign conducted in July 2012. Finally, in Section 7 we summarize the plan for some possible future activities on other type of X-ray instruments as well.

## 2. SILICON DRIFT DETECTOR AND LOFT MISSION

**2.1 Silicon Drift Detector**

We give here an essential introduction to the *Silicon Drift Detector* (SDD) concept [1],[2]. SDD is a type of photo-diode, functionally similar to a PIN photo-diode, but with a unique electrode structure to improve performance. SDD consists of a volume of fully depleted high-resistivity Silicon, in which an electric field with a strong component parallel to the surface drives the electrons generated by the absorption of ionising radiation towards a set of small-sized collecting anodes. The electric field is generated by a series of cathodes placed on both surfaces, polarized with a negative voltage progressively decreasing fom the central cathode down to the anodes. The key advantage of the SDD is that, thanks to the small area of the read-out anodes, it has much lower capacitance than a conventional diode. Therefore electronic noise at short shaping times is reduced and, for X-ray spectroscopy, the SDD has better energy resolution than a diode while still operating at high count rates.

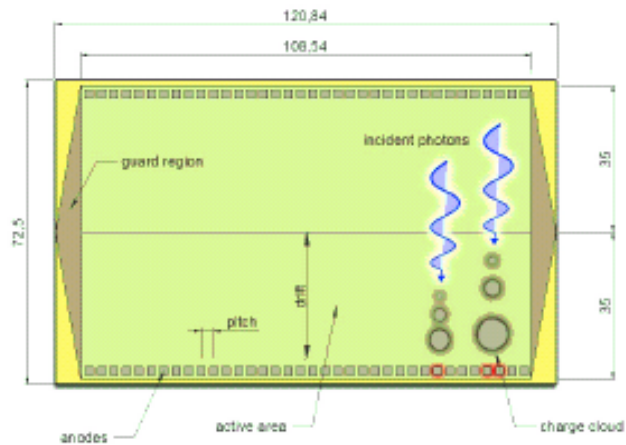

Figure 1. *LOFT* Silicon Drift Detector scheme

The type of SDD used on *LOFT* is a version derived from that developed by INFN-Trieste & Canberra Inc. [3] for the

ALICE experiment at CERN/LHC. It is a large area (about 11 cm x 7 cm) Silicon tile with 450 μm thick depletion volume coated by a superficial oxide structure (Figure 1). A drift time as low as ~ 7 μsec is achieved. The leakage current is low enough to allow for the SDD being operated even at room temperature. However, the radiation environment in space can cause a severe increase of the leakage current. Such effect can be partially compensated by a moderate (-20/-30 °C) cooling in orbit.

## 2.2 LOFT mission

*LOFT (Large Observatory for X-ray Timing)* is a newly proposed M-class space mission devoted to perform fast X-ray timing measurements in the energy range 2-50 keV by exploiting a very large collecting area [4]. It has been selected in 2011 by ESA as a possible candidate mission for a launch in 2020's and the project is currently in the assessment phase. The mission design (Figure 2) prefigures two instruments both based on SDDs, the *Large Area Detector* (LAD) and the *Wide Field Monitor* (WFM). LAD (Figure 3) is a collimated instrument that consists of 6 deployable panels, each panel is composed by 21 modules, each module contains 4x4 SDDs. The total geometric area is ~18 m$^2$. WFM (Figure 4) is a coded aperture imaging instrument consisting of 4 units, each unit is composed by 2 co-aligned SDD cameras. The difference between LAD SDDs and WFM SDDs is the anode pitch, 970 μm and 145 μm respectively. A more detailed description of *LOFT*, LAD and WFM can be found in [4],[5],[6] .

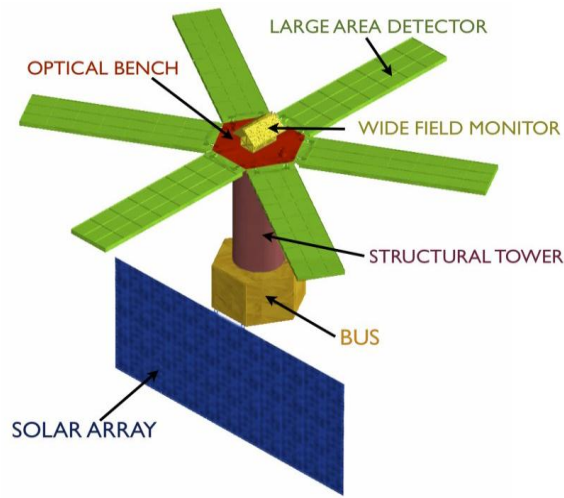

Figure 2. LOFT conceptual scheme (courtesy of ISDC)

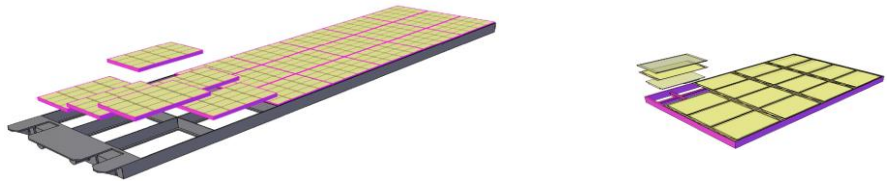

Figure 3. LAD: single panel (*left*) and single module (*right*)

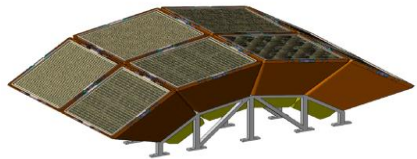

Figure 4. WFM: 8 co-aligned cameras (4 units)

# 3. LOFT SPACE ENVIRONMENT: RADIATION AND DUST

## 3.1 Radiation environment: soft proton flux

The current baseline prefigures for *LOFT* an equatorial Low Earth Orbit at ~600 km altitude with inclination 5º or less. The space environment in this orbit presents the advantage of solar and cosmic particle fluxes largely reduced by the geo-magnetic field, and a minimization of the effects related to South Atlantic Anomaly (SAA) crossings. In particular, in a LEO the softer (less than a few GeV, on average) cosmic component of radiation is completely cut away, as well as solar particles emitted in form of wind and impulsive events. Nevertheless, there are albedo particles (charged particles, neutrons and gamma-rays) generated by the interactions of high energy cosmic rays with the atmosphere, that also can affect the performance of a detector. Between ~1 MeV and ~1 GeV albedo protons are the main component of charged radiation. However, in an equatorial orbit the proton flux below ~1 MeV is enhanced by a component originated from the second stage charge exchange process [6]. The flux of these very soft protons is found to be almost independent of the altitude between 500 and 1000 km and can be described by a kappa-function (Figure 5) [7], with some dependence on geo-magnetic activity and pitch-angle (peak at ~90 deg).

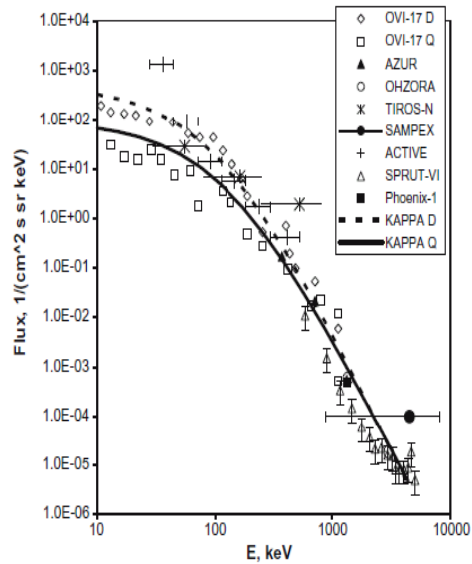

Figure 5. Near-equatorial proton spectrum [7]

## 3.2 Dust environment: micro-meteoroids and debris populations

The Earth is surrounded by a belt of micro-meteoroids and orbital debris (Figure 6), made of natural fragments and remainders of older and decommissioned spacecraft, that are potentially dangerous for satellites in orbit. *LOFT*, due to its light structure and small mass per unit surface, may be particularly sensitive to this type of hazard. As a part of the *LOFT* SDDs qualification process, we are performing a study, through theoretical calculations, simulations and laboratory tests, aimed to assess the potential risk of impacts on the SDDs by micro-meteoroids and debris. However, the risk posed by micro-meteoroids is expected to be lower than that associated with the debris. The mass of natural fragments with size less than 1 mm is estimated to be about 200 kg within 2000 km above the Earth's surface, while within this same distance the estimated mass of debris with size less than 1 mm is about 300 kg. While the flux of micro-meteoroids is isotropic, the flux of orbital debris can be higly directional, with most of debris found in high inclination orbits. Debris can be divided into six groups based on particle size. The smallest size group (with particle diameter less than 20 μm) is associated with the highest flux, which makes this type of particles the most dangerous ones. They are $Al_2O_3$ grains generated as the result of solid rocket motor burns [8].

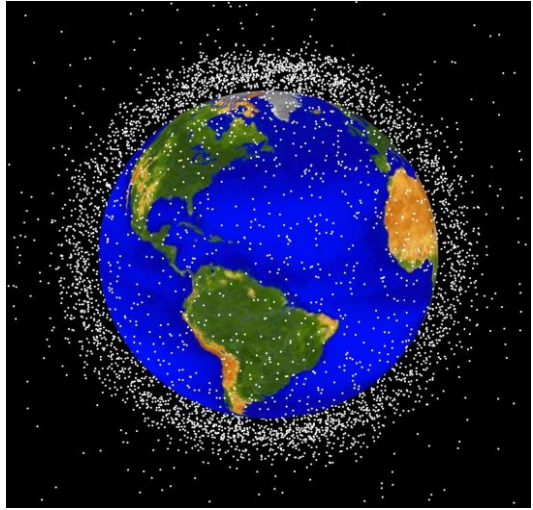

Figure 6. Orbital debris around the Earth

## 4. EXPECTED EFFECTS OF SOFT PROTON BOMBARDMENT AND HYPER-VELOCITY PARTICLE IMPACTS

**4.1 Consequences of soft proton irradiation**

Protons impinging onto the SDDs cause ionization and irreversible crystallographic defects, which can be electrically active and lead to variations in the space charge, increase of leakage current and trapping. Soft protons can typically create point defects. The simplest defect in Silicon is the so-called *Frenkel pair*, in which a Silicon atom is displaced from its lattice site to the interstitial. The two expected consequences of proton irradiation are a degradation of the field oxide layer and of the interface $SiO_2$-Si due to ionization that can lead to an increase of the surface leakage current, and a displacement damage in the bulk, that will determine an increase of the bulk leakage current. The total leakage current is the sum of the surface and bulk contributions. Since protons below ~1 MeV can penetrate only a few tens of microns below the surface (Figure 7), they are expected to possibly affect the surface leakage current, while protons above ~ 1MeV would produce effects in the bulk. In order to qualify the *LOFT* SDDs the effects on softer and harder protons are investigated separately. The SDDs performance under proton irradiation in the bulk is tested at the facility of the Paul Scherrer Institute, with protons in the range 1-50 MeV. The SDDs under proton irradiation on the surface are tested at the facility of the University of Tübingen with protons in the range 0.1-1 MeV, as described in Section 5. From the flux reported in [7] we calculate the dose of soft protons ~0.1-1 MeV during the nominal duration of *LOFT* mission (~0.1 MeV is a lower threshold related to the presence of filters in front of the SDDs, 80 nm Al + 1000 nm Kapton for LAD, 100 nm Al + 6700 nm Kapton + 160 nm $SiO_2$ + 25000 nm Be for WFM). Then, the expected nominal values of *total ionizing dose* (TID) and leakage current increase from *non-ionizing energy loss* (NIEL) are analytically computed for LAD and WFM [9]. Predicted TID values after 5 years in orbit are ~70 rad($SiO_2$) for LAD and ~155 rad ($SiO_2$) for WFM; predicted NIEL induced leakage current increases are ~32.5 pA/anode for LAD and ~36.5 pA/anode for WFM. Values are different for LAD and WFM due to the different anode structure, filter layers and field of view (~1º x 1º for LAD, ~180º x 90º for WFM). From these numbers, the total proton dose is converted into two damage-equivalent doses of 200 keV and 800 keV protons, in order being able to experimentally reproduce the effects of both TID and NIEL and validate expectations. We obtain equivalent doses of ~7·$10^6$ p/$cm^2$ @ 200 keV and of ~1.6·$10^6$ p/$cm^2$ @800 keV for LAD, and of ~3.55·$10^6$ p/$cm^2$ @200 keV and of ~3.42·$10^7$ p/$cm^2$ @800 keV for WFM [9]. These values are reported also in Table 1.

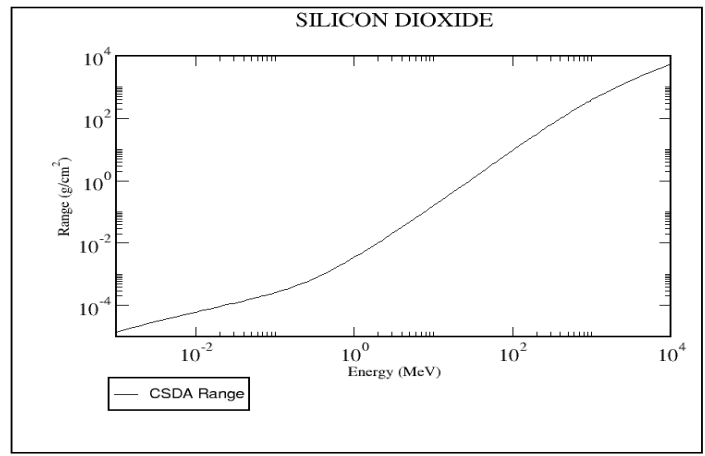

Figure 7. Range of protons in $SiO_2$ (from NIST)

**4.2 Consequences of hyper-velocity particle impacts**

The impact of micro-meteoroids and debris will cause craters on the surface of the target. The aim of our tests is to determine to which extent the formation of craters may degrade the SDDs performance. A possible significant increase of the SDD leakage current and of the drift field may be expected. However, such effects have never been explored up to now and therefore there is a lack of reference literature on this topic. A preliminary estimation of the expected rates in orbit has been calculated using the ORDEM2000 software package. Based on this, we expect on LAD an impact-rate ~10 impact/year by 1μm size projectiles and ~1 impact/year by 10 μm size projectiles. The impact-rates expected for WFM are ~600 impact/year by 1μm size projectiles and ~60 impact/year by 10 μm size projectiles, these rates are larger than LAD values due to the much larger field of view of WFM. However, ORDEM2000 is a quite old model and we will then refine the analysis by dedicated simulations with the ESABASE2 toolkit, which exploits updated debris datasets and models and allows for a more complete analysis of the particle impacts. Then, we put this values in relation with the impact particle parameters to assess the risk of failure. Grun [10] developed an equation to approximate the estimate of the depth of the craters produced by the impact of micron-sized Al and Fe particles:

$$P = 0.772 * d^{1.2} * (v * \cos(\theta))^{0.88} * \rho_1^{0.73}/(\varepsilon^{0.06} * \rho_2^{0.5})$$

where P is the crater depth (cm), d (cm) is the diameter of the impacting particle, v is the speed of the impacting particle (km/sec), θ is the impact angle, ε is a ductility constant, $\rho_1$ (g/cm$^3$) and $\rho_2$ (g/cm$^3$) are the densities of the impact particle and target respectively. Since the expected average debris velocity is ~ 10 km/sec [11], from this equation we expect the production of craters with a few microns depth may on the SDDs surface.

## 5. PRELIMINARY ACCELERATOR TEST WITH SOFT PROTONS

**5.1 Experimental set-up: Van de Graaff accelerator in Tübingen**

To investigate the effects of soft protons on the SDDs we use the Van de Graaff accelerator of the University of Tübingen. This accelerator can provide a proton beam with energy up to ~3.5 MeV. Currently, the maximum energy available is 2.3 MeV. We performed a first test on a SDD prototype in June 2012. As explained in Section 4 /4.1, we wanted to operate with two quasi-monoenergetic beams @ ~200 keV and @ ~800 keV, having spatial extent large enough to allow for a uniform irradiation of the SDD surface. The desired properties have been obtained by the use of degrader foils. With the TRIM simulator we calculated the optimal materials and thicknesses to be interposed between the proton beams and the SDD [12]. We used a 6 μm Cu foil to obtain a quasi-monoenergetic (FWHM ~ 70 keV) beam at ~200 keV, starting from 1 MeV energy from the accelerator. We used an 18 μm Cu foil to obtain a quasi-monoenergetic (FWHM ~100 keV) beam at ~800 keV, starting from 2.3 MeV from the accelerator. The degrader foils also introduced lateral straggling of the beams that provided the spatial broadening (about ±20 deg) needed for a nearly uniform irradiation over the whole SDD surface (that in the adopted beamline geometry sees only protons incoming from about ±1 deg) [12]. We installed a SDD prototype (Figure 8 and Figure 9) provided by INFN-Trieste inside the chamber, in vacumm at ~$10^{-5}$ mbar, together with four surface-barrier proton counters at the sides of the SDD to monitor the flux and its uniformity. The SDD was supported by a special PCB frame and was divided into two halves having anode pitches 835 μm (LAD side) and 294 μm (WFM side). At the four corners of the SDD there were two MOS structures and two gated diodes (GD#2 and GD#3) having the same structure of the SDD, working as characterization/control devices. We irradiated the whole SDD surface in four irradiation steps with the doses of protons reported in Table 1[8]. During the irradiations both the SDD and the gated diodes were biased at 30 V.

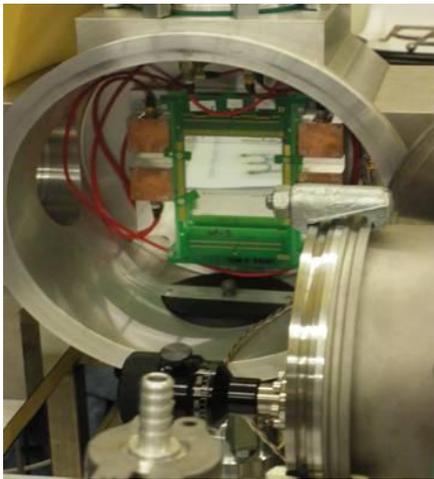

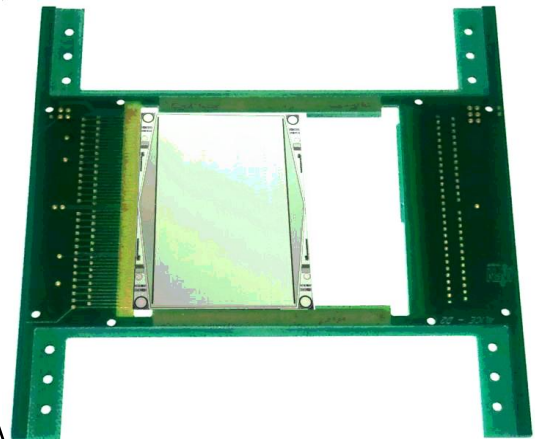

Figure 8. SDD+PCB prototype for soft proton test

Table 1. Irradiation steps @ 200 keV and @800 keV[8]

| | $LAD_{200}$ | $LAD_{800}$ | $WFM_{200}$ | $WFM_{800}$ |
|---|---|---|---|---|
| #1 | 50% d | 50% d | 98.6% d | 3.3% d |
| #2 | 100% d | 100% d | 197.2% d | 6.6% d |
| #3 | 100% d | 1500% d | 197.2% d | 70% d |
| #4 | 101% d | 4100% d | 200% d | 200% d |
| | 4.5x | 15y | 2x | 2y |

**d = dose expected after 5 years in orbit** [8]
d ~ $7 \cdot 10^6$ p/cm$^2$ @200 keV (LAD)
d ~ $1.6 \cdot 10^6$ p/cm$^2$ @800 keV (LAD)
d ~ $3.55 \cdot 10^6$ p/cm$^2$ @200 keV (WFM)
d ~ $3.42 \cdot 10^7$ p/cm$^2$ @800 keV (WFM)

**x = nominal TID    y = nominal NIEL**
$x_{LAD}$ ~69 rad(SiO$_2$)   $y_{LAD}$ ~32.5 pA/anode
$x_{WFM}$ ~155 rad(SiO$_2$)  $y_{WFM}$ ~36.5 pA/anode

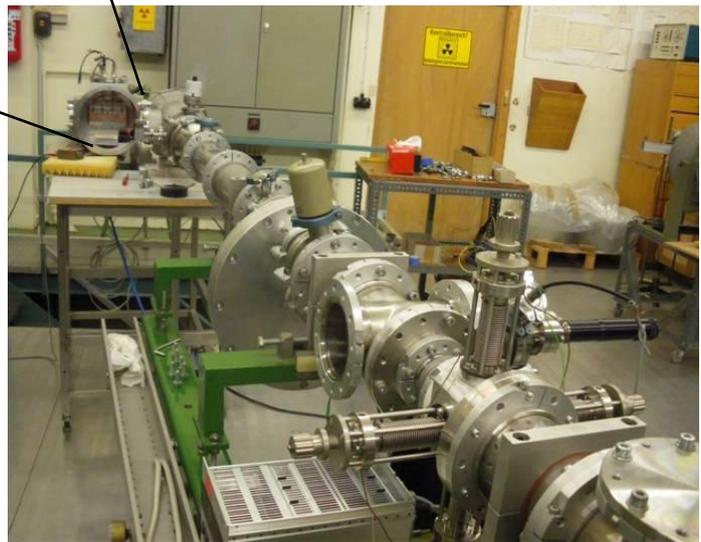

Figure 9. Beamline set-up at the Van de Graaff accelerator (zoom on the test chamber with the mounted SDD and proton counters)

## 5.2 Some preliminary data analysis and interpretation

Table 1 reports the proton doses accumulated after each step of irradiation for LAD and WFM. At the end of step #4 the doses are such that the corresponding TID and NIEL amounts for WFM are ~2 times the nominal expected ones, while those for LAD are ~4.5 times (TID) and ~15 (NIEL) times the nominal expected ones. Notice that due to the different anode pitches of the tested prototype with respect to the baseline design, the leakage current increase from NIEL in the bulk should be ~2 times larger than the actual one for the WFM anodes, while for the LAD anodes it should be slightly lower. The increase in NIEL leakage current for the gated diodes is expected to be ~36 pA/diode. We monitored the I-V behaviour of the gated diodes GD#2 and GD#3 in between irradiation steps. GD#2 showed (even before the first irradiation step) some extra contribution superimposed to the expected curve, possibly generated by the presence of a dust grain or grease on the connectors. GD#3 showed the expected I-V profile. We report (Figure 10) a comparison of the I-V curves of GD#3 measured at INFN-Trieste before shipping the SDD to Tübingen and after receiving back the irradiated SDD (four days after finishing the irradiation campaign). The gate polarization was scanned from -5 V to +10 V, in such a way that all the states (inversion, depletion, accumulation) of the diode are visible. The surface leakage current is measured from the amplitude of the transition inversion - depletion. An increase of the surface leakage current may be mainly attributed to TID effects. An increase of the offset current in inversion regime gives indication of NIEL damage. The measured increase of the surface leakage current was ~2.5 pA/mm$^2$. The measured increase in bulk leakage current (inversion offset) was ~ 2 pA/diode. Both values are much less than expected based on the annealing effet. After one week, the I-V curve recovered completely to the initial profile. A complete analysis on the SDD anode leakage current is presently ongoing at INFN-Trieste.

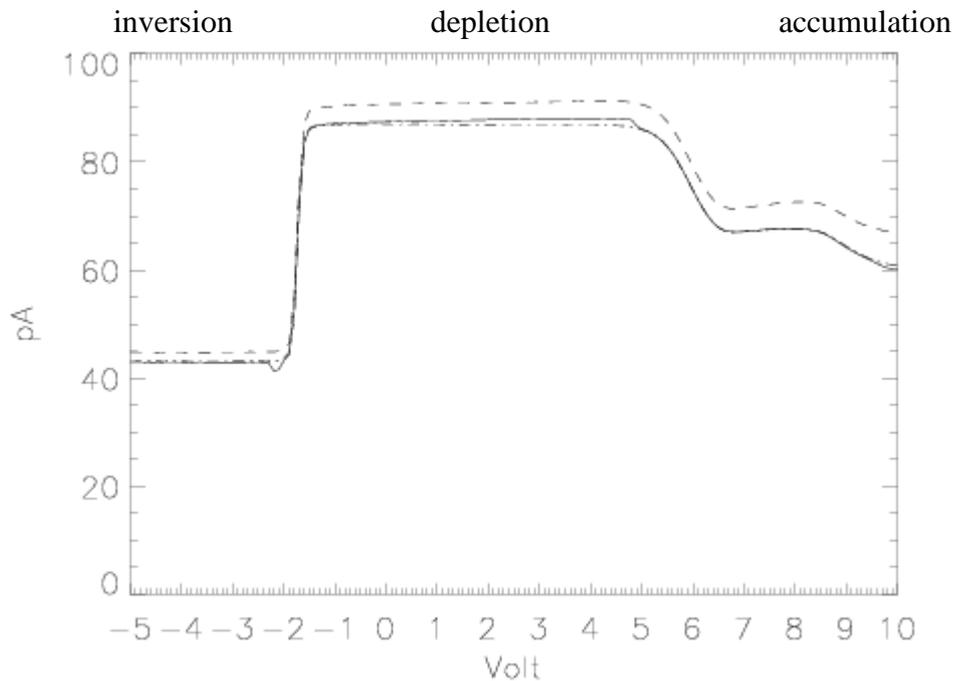

Figure 10. I-V curve for GD#3 measured at INFN-Trieste before and after proton irradiation (*dotted* line=4 days after irradiation, *dotted-dashed* line=11 days after irradiation)

# 6. PRELIMINARY ACCELERATOR TEST WITH DEBRIS

**6.1 Experimental set-up: Van de Graaff accelerator in Heidelberg**

A preliminary test with debris on SDD has been performed in July 2012 at the facility of the Max Planck Institute for Nuclear Physics (MPIK) in Heidelberg. The dust accelerator is a modified Van de Graaff generator, able to reach a potential of 2 MV. A complete description of the accelerator can be found in [13]. The accelerator (Figure 11 and Figure 12) can be operated both in *single-shot* and in continuous mode. Since we expect a relatively low impact-rate in orbit, the first option, which allows to select and change during the experiment the main parameters (size,speed) of each particle in each shot, resulted preferable. An SDD prototype similar to that shown in Section 5 was installed inside the big (1.4 m diameter) chamber available at MPIK, in vacuum at $2.5 \cdot 10^{-6}$ mbar (Figure 13).

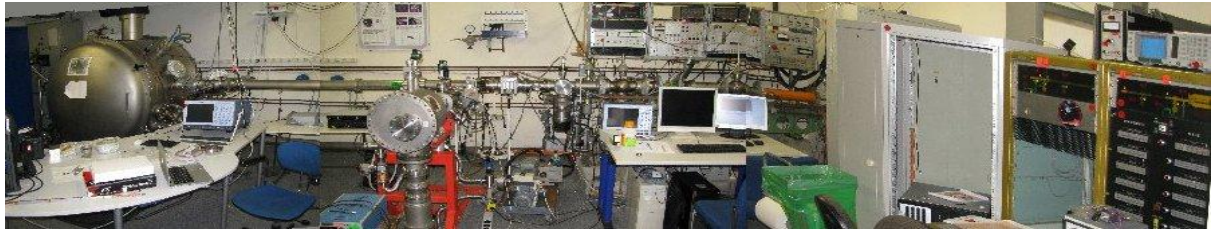

Figure 11. Beamline set-up of the Van de Graaff dust accelerator at MPIK

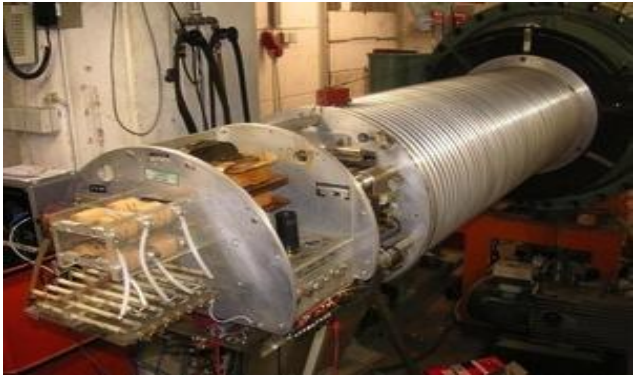

Figure 12. Accelerating coils

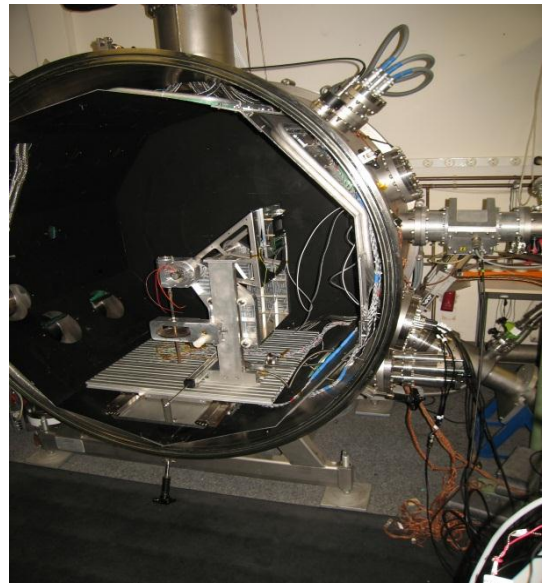

Figure 13. The SDD installed in the big test chamber

The mounting inside the chamber permitted to move manually the detector along the x-y directions, and to tilt it with respect to the direction of incident particles. An encoder resistance was used to select the position along the x-axis. The set-up included a power supply to bias the SDD at its maximum voltage of -1300 V, and two *Keithley* electrometers interfaced with a computer through a *Velleman* data logger, to monitor and record the total anode leakage currents from the two halves of the SDD. All the anodes of the SDD prototype selected for this test have been bonded together in order to obtain a higher leakage current signal. Two triangular regions of the sides of the active area of SDD, where the guard electrodes have been implemented, are expected to be particularly sensitive to debris impacts: hitting them could be critical. In LOFT the LAD/SSDs guard regions can be shielded by properly shaping the collimators, while the

WFM/SDDs will be protected partially by the same filtering mechanism that on the rest of the detector. Therefore, we decided to start the test with an Al mask covering the guard regions and anodes, simulating the likely actual LAD configuration. We shot several debris particles made of *olivine* with size in the range 0.1-2.5 μm and velocity in the range 0.5-34 km/sec (Figure 14 shows the trade-off size-velocity available at the accelerator) onto the active area of the SDD, on both the LAD and WFM halves. The precision in hitting was within about 1 cm$^2$, and could be directly controlled with a laser beam coaligned with the collimator. A charge-detector placed at the entrance of the test chamber provided an alert each time a particle entered the chamber and hit the SDD.

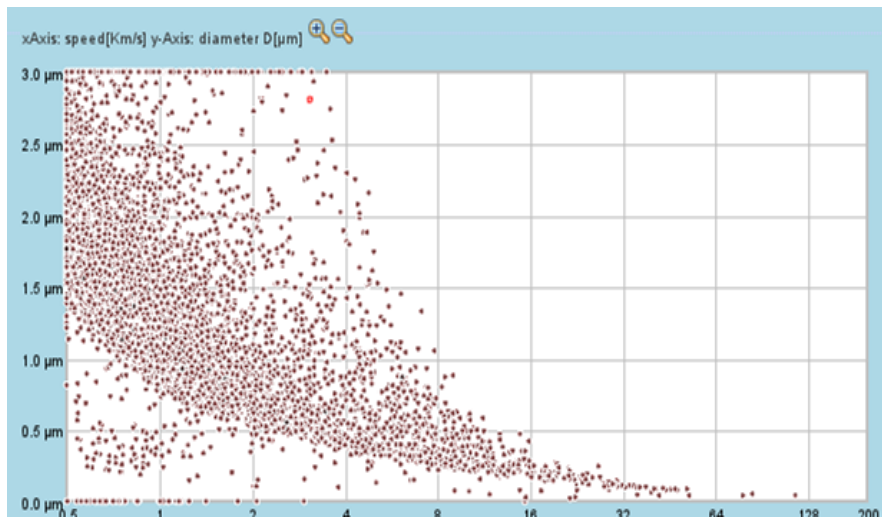

Figure 14. Distribution size-velocity of the *olivine* debris particles generated by the accelerator

**6.2 Some preliminary data analysis and interpretation**

A detailed analysis of the post-test condition of the SDD is presently ongoing at INFN-Trieste. During the bombardment at MPIK we monitored in real-time the behaviour of the SDD, and then we observed it at the microscope to see craters produced by the impacts (Figure 15). This was a preliminary test, and the SSD survived to debris bombardment. Also, the degradation of the performance seems less than expected.However, for the time being we prefer do not anticipate here other details on the results of the test since they are currently matter of further investigation,for example to understand how many of the particles we shot were actually able to penetrate into the SDD structure deeply enough to produce a possible damage. We do this by means of simulations (ESABASE2) and available empirical models.

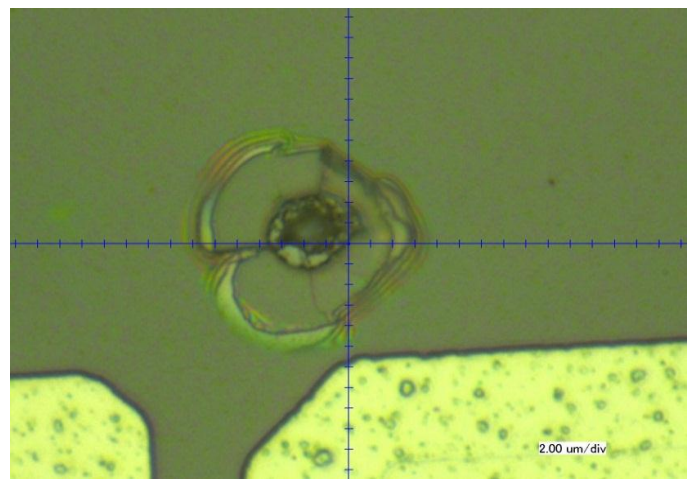

Figure 15. Microscope view of a crater produced in the voltage divider area by a debris impact

## 7. FUTURE WORK

Besides other more complete tests on *LOFT* SDDs, future accelerator activities are programmed with other type of X-ray instrumentation as well. In the following we summarize some of them:

### 7.1 Tests on micro-calorimeters

The *ASTRO-H* mission [14], which is scheduled for launch in 2014, will be equipped with an array of X-ray micro-calorimeters. The very thin structure (a few micron thickness) of this detector and related wiring and the evidence that micro-meteoroids and debris can be scattered off X-ray mirror shells down to the focal plane suggest to perform some tests to verify how a micro-calorimeter can withstand hyper-velocity dust impacts. For these purposes we are planning to test a few samples provided by the group at NASA –Goddard Space Flight Center (GSFC) using a small chamber available at MPIK where a micro-calorimeter detector box can be mounted below a 6 mm diameter collimator (the set-up is showed in Figure 16).

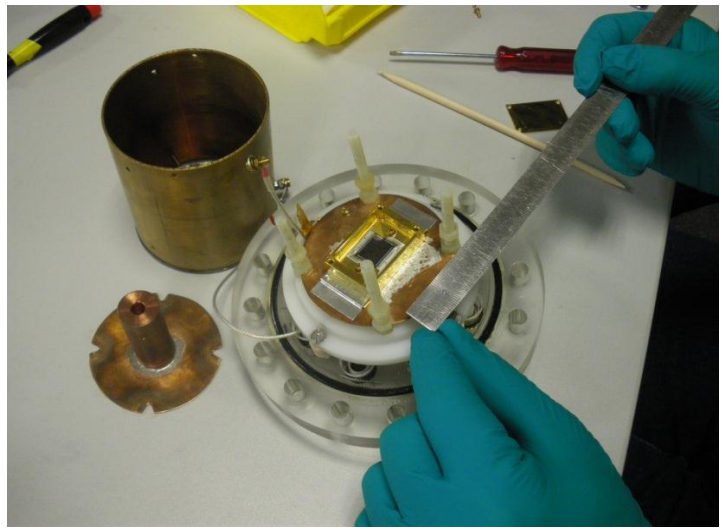

Figure 16. A 6x6 array of GSFC micro-calorimeters mounted in the small test chamber at MPIK for debris impact test

### 7.2 Tests on pn-CCD

The SVOM mission [15] will use a pn-CCD at the focal plane of the MXT telescope. Since SVOM is planned in a LEO orbit with ~30º inclination and the detector cannot be cooled below -60 °C, the proton fluence during SAA crossings poses a high risk for this detector. This type of pn-CCD is also used aboard the eROSITA mission and its robustness and performance degradation under proton irradiation has already been extensively tested up to doses ~$5.6 \cdot 10^8$ p/cm$^2$ [16]. However, the dose expected at the end of the SVOM mission is almost two times larger than this. Therefore, some new tests at higher doses are mandatory.

## 7.3 Tests on X-ray mirrors

The evidence [17],[18] that X-ray mirror shells can scatter soft protons and dust particles down to focal plane detector. stresses the need of dedicated accelerator tests to improve and validate the existing models for the assessment of the risk posed by scattering from mirrors. We plan to perform some scattering experiments using as target some parts of eROSITA mirror shells cut from a spare telescope prototype at MPE (Figure 17). They are a few centimeters in size so that they can easily fit inside the available test chambers.

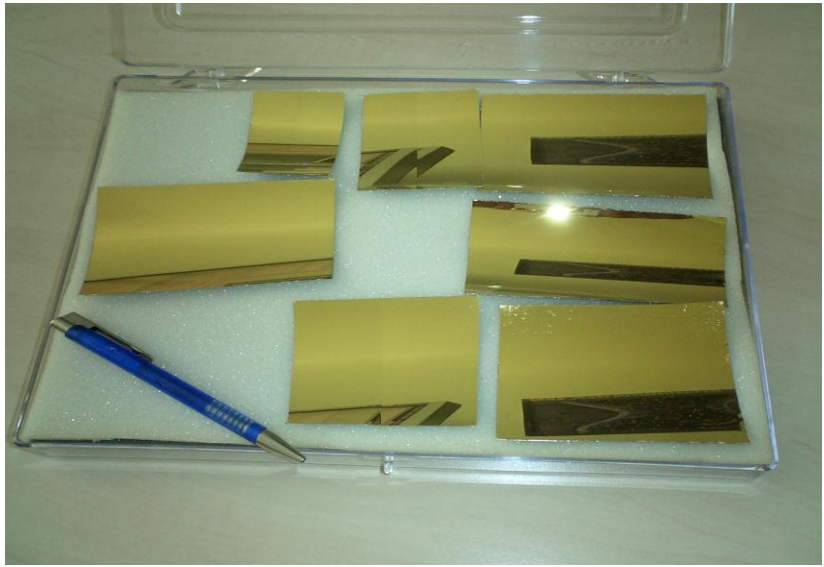

Figure 17. Some mirror samples cut from a spare eROSITA shell at MPE

## 8. CONCLUSIONS

We presented an overview on experimental activities with particle accelerators turned to characterize the performance of X-ray instrumentation for future X-ray missions. These activities are carried out within a large collaboration involving different groups on different projects, with the aim to optimize time, efforts and costs in the use of accelerator facilities. We performed a first irradiation test with soft protons in June 2012 on a SDD prototype for LOFT. A first tests with hyper-velocity dust particles has been performed in July 2012. Both tests demonstrated the robustness of the SDD structure and a degradation of its performance less than expected. More tests on SDDs and other type of X-ray instrumentation, such as micro-calorimeters, are in progress or planned in the nearest future.

## ACKNOWLEDGEMENTS

This work is partially supported by the *Bundesministerium für Wirtschaft und Technologie* through the *Deutsches Zentrum für Luft und Raumfahrt* (Grant FKZ 50 00 1110). Authors wish to acknowledge all other institutions and agencies supporting these activities. In particular, J.P. Osborne acknowledges financial support from the UK Space Agency. We are grateful to the Physikalisches Institut Tübingen that made available the accelerator for this research program, and to MPIK for providing timely a slot for tests .

# REFERENCES

[1] Gatti, E. Et al, "*Semiconductor Drift Chamber-an application of a novel charge transport scheme*", Nuclear Instruments and Methods A **225**, pp. 608-614,1984

[2] Lechner, P. et al., "*Silicon Drift Detectors for high resolution room temperature X-ray spectroscopy*", Nuclear Instruments and Methods A **377**, pp. 346-351,1996

[3] Rachevski, A. et al., "*Large area Silicon Drift Detector for the ALICE experiment*",Nuclear Instruments and Methods A **485**,pp. 54-60,2002

[4] Feroci, M.et al., "*The Large Observatory for X-ray Timing (LOFT)*", Exp. Astr., ?, pp.1-30,2011

[5] Zane, S. et al., "*A Large Area Detector proposed for LOFT*" , these SPIE Proceedings

[6] Brand, S. et al., "*The LOFT Wide Field Monitor*", these SPIE Proceedings

[7] Petrov, A N. et al. "*Creation of model of quasi-trapped proton fluxes below Earth's radiation belts*", Advances in Space Research **43**, pp. 654-658,2007

[8] Kessler, D. J. et al., "*A computer-based orbital debris environment model for spacecraft design and observation in Low Earth Orbit*", NASA TM-104825

[9] Zampa, N. et al., Technical Report (in prep.)

[10] Grun, E. et al.,"*The penetration limit of thin films*", Planetary Space Science **28**, pp. 321-331,1979

[11] Anderson, B.J.et al, "*Natural orbital environment guidelines for use in aerospace vehicle development*", NASA TM 4527

[12] Diebold, S. et al, Technical Report (in prep.)

[13] Mocker, A. et al., "*A 2 MV Van de Graaff accelerator as a tool for planetary and impact physics research*", Review of Scientific Instruments **82**,pp. 095111, 2011

[14] Takahashi,T. et al., "*The ASTRO-H mission*", SPIE Proceedings **7732**,pp. Z1-Z18,2010

[15] Götz,D. et al., "*SVOM:a new mission for Gamma-Ray-Bursts studies*", AIP Proceedings 1123,pp. 25-30,2009

[16] Meidinger, N.et al., "*Development for the eROSITA Space Telescope*", IEEE Nuclear Science Symposium Conference Record **N02-1**,pp.24-31,2010

[17] Strüder et al.,"*Evidence for micrometeoroid damage in the pn-CCD camera system aboard XMM-Newton*", A&A **375**, pp. L5-L8,2001

[18] Carpenter, J. D. et al., "*Meteoroid and space debris impacts in grazing incidence telescopes*", A&A **483**,pp.941-947,2008